\documentclass[twocolumn,prb,color,superscriptaddress,psfig,showpacs,amsmath,amssymb,floatfix,fleqn]{revtex4-1}

\usepackage{natbib}
\setcitestyle{square,numbers}

\usepackage{amsmath}
\usepackage{amsfonts}
\usepackage{amssymb}
\usepackage{amsthm}
\usepackage{mathtools}

\usepackage{epsf}
\usepackage{graphicx}
\usepackage{epstopdf}

\usepackage{color}
\newcommand{\vl}[1]{\textcolor{black}{#1}}

\begin{document}

%\title{Magnetic properties of the spin-1 chain compound NiCl$_3$C$_6$H$_5$CH$_2$CH$_2$NH$_3$ as probed by ESR and NMR spectroscopies}

\title{Magnetic properties of the spin-1 chain compound NiCl$_3$C$_6$H$_5$CH$_2$CH$_2$NH$_3$}

\author{F.~Lipps}
\affiliation{Leibniz Institute for Solid State and Materials Research IFW Dresden, D-01171 Dresden, Germany}

\author{A. H. Arkenbout}
\affiliation{Zernike Institute for Advanced Materials, University of Groningen, Nijenborgh 4, 9747 AG Groningen, The Netherlands}

\author{\vl{A. Polyakov}}
\affiliation{Zernike Institute for Advanced Materials, University of Groningen, Nijenborgh 4, 9747 AG Groningen, The Netherlands}

\author{M. G\"{u}nther}
\affiliation{Institut f\"{u}r Festk\"{o}rperphysik, TU Dresden, D-01069 Dresden, Germany}

\author{\vl{T. Salikhov}}
\affiliation{Kazan E. K. Zavoisky Physical Technical Institute of RAS, 420029 Kazan, Russia}

\author{E. Vavilova}
\affiliation{Kazan E. K. Zavoisky Physical Technical Institute of RAS, 420029 Kazan, Russia}

\author{H.-H. Klauss}
\affiliation{Institut f\"{u}r Festk\"{o}rperphysik, TU Dresden, D-01069 Dresden, Germany}

\author{B.~B\"{u}chner}
\affiliation{Leibniz Institute for Solid State and Materials Research IFW Dresden, D-01171 Dresden, Germany} \affiliation{Institut f\"{u}r
Festk\"{o}rperphysik, TU Dresden, D-01069 Dresden, Germany}

\author{T. M. Palstra}
\affiliation{Zernike Institute for Advanced Materials, University of Groningen, Nijenborgh 4, 9747 AG Groningen, The Netherlands}

\author{V.~Kataev}
\affiliation{Leibniz Institute for Solid State and Materials Research IFW Dresden, D-01171 Dresden, Germany}

\date{\today}

\begin{abstract}

We report experimental results of the static magnetization, ESR and NMR spectroscopic measurements of the Ni-hybrid compound
NiCl$_3$C$_6$H$_5$CH$_2$CH$_2$NH$_3$. In this material NiCl$_3$ octahedra are structurally arranged in chains along the crystallographic $a$-axis.
According to the static susceptibility and ESR data Ni$^{2+}$ spins $S = 1$ are isotropic and are coupled antiferromagnetically (AFM) along the chain
with the exchange constant $J = 25.5$\,K. These are important prerequisites for the realization of the so-called Haldane spin-1 chain with the
spin-singlet ground state and a quantum spin gap. However, experimental results evidence AFM order at $T_{\rm N} \approx 10$\,K presumably due to
small interchain couplings. Interestingly, frequency-, magnetic field-, and temperature-dependent ESR  measurements, \vl{as well as the NMR data},
reveal signatures which could presumably indicate an inhomogeneous ground state of co-existent mesoscopically spatially separated AFM ordered and
spin-singlet state regions similar to the situation observed before in some spin-diluted Haldane magnets.

\end{abstract}


%\keywords{superconductor,ferromagnet,proximity effect}

\maketitle
\section{Introduction}

Investigations of quantum magnetic phenomena in spin networks with reduced spatial dimensions of magnetic interactions is a well established and
exciting field of research in condensed matter physics (for reviews see, e.g., Refs.~\cite{Jongh1990,Mikeska04,Sachdev08,Zvyagin10}). In systems with
reduced dimensionality quantum effects become more relevant and ground states can be established not observed in three-dimensional systems. Ground
state properties and excitation spectra depend critically on the dimensionality of the interaction, the dimensionality of the spin and the interplay
between different interactions. On the experimental side, the search for realizations of the spin systems where magnetic exchange between the
localized spins is restricted to one (1D) or two (2D) spatial dimensions is important for the verification of modern theories of quantum magnetism
and for the exploration of novel magnetic phenomena.

Indeed, in many naturally occurring or man made solids the magnetic interactions are restricted to less than their three dimensions. This is the case
when the crystal structure assembles in such a way that the couplings between spins along certain directions are much stronger than along others.
There are 2D systems in which interaction takes place predominantly between magnetic ions arranged in a plane. In other systems magnetic ions are
arranged in 1D structures, forming so-called spin chains.

One of the important classes of spin chains is the Haldane chain. This is a one-dimensional Heisenberg chain with integer spins and antiferromagnetic
(AFM) nearest-neighbor coupling. Haldane predicted that the ground state of such a system would be a non-magnetic singlet state which would be
separated in energy from the excited triplet state by a gap $\Delta$ \cite{Haldane1983}. This gap is not an anisotropy gap, but is due to the quantum
nature of the $S=1$ system. Haldane considered the pure Heisenberg Hamiltonian for an easy axis configuration \cite{Haldane1983}. In order to explore
the limits of the Haldane phase bi-quadratic exchange and single ion-anisotropy, among other parameters, can be taken into account. Already those
simple extensions reveal rich physics involved in the quasi-one-dimensional antiferromagnetic integer Heisenberg spin chains. A general Hamiltonian
of the Haldane system is given in \cite{Renard2001}:
\begin{eqnarray}
\mathcal{H} & =  & J\sum_{i}[\mathbf{S}_{i}\mathbf{S}_{i+1}  +  \beta(\mathbf{S}_{i}\mathbf{S}_{i+1})^2] \nonumber \\
& + & \sum_{i}[D(S_{i}^{z})^2-g\mu_{\rm B}S_{i}^{\alpha}H^{\alpha}]. \label{Hamilton}
\end{eqnarray}
Here $J$ is the energy coupling constant between neighboring spins $\mathbf{S}$. $\beta$ describes the bi-quadratic exchange. Uniaxial single ion
anisotropy is considered: With $z$ being the chain direction either easy-axis (spin along the chain, $D<0$) or easy-plane (spin perpendicular to the
chain, $D>0$) is favored. The interaction with a magnetic field $H$ is described by the typical Zeeman term where $g$ is the $g$-factor and $\mu_{\rm
B}$ is the Bohr magneton.

%A good visualization of the Haldane system is the so-called valence-bond solid (VBS). Here two spins on different ions with opposite spins are
%thought to couple strongly and form a non-magnetic singlet. The VBS state was calculated as a special realization of the Haldane phase considering
%bi-quadratic exchange \cite{Affleck1987}, but physics for VBS and the pure Heisenberg hamiltonian are found to be identical \cite{Renard2001}. For
%larger contributions of bi-quadratic exchange other ground states are realized. Dimerized and trimerized antiferromagnetic ground states and a
%ferromagnetic phase have been identified \cite{Renard2001}.\\
%
Besides the bi-quadratic exchange the single ion anisotropy plays a crucial role for the realization of a Haldane system. The energy gap $\Delta$
between the singlet ground state $|0\rangle$ and the excited triplet state $|1\rangle$ directly depends on the value of $D$. The gap is largest for
the absence of single ion anisotropy, but an energy difference exists within a certain range of $D$. For $D >J$ an anisotropy gap opens.

The first material discovered to realize the Haldane system was Ni(C$_2$H$_8$N$_2$)$_2$-NO$_2$ClO$_4$ (NENP) \cite{Renard1987}. Similar to the system
studied in this present work NENP contains the transition element Ni realizing the chain structure in an organic matrix. NENP exhibits a single ion
anisotropy \cite{Renard1987}, which results in the splitting of the excited triplet state \cite{Sieling1995} and with that an anisotropic Haldane
gap. However the ground state is still the singlet state. Since then a number of other compounds featuring Ni-based chains have been discovered and
investigated (for details see, e.g., Refs.~\cite{Renard2001,Wierschem14}).

In the present paper which summarizes some of the results of the PhD work in Ref.~\cite{LippsThesis2011} we report the magnetic properties of the
inorganic-organic hybrid compound with the chemical formula NiCl$_3$C$_6$H$_5$CH$_2$CH$_2$NH$_3$ as revealed by static magnetic measurements, and ESR
and NMR local probe techniques. This compound contains structurally well isolated Ni chains where Ni$^{2+}$ spins $S = 1$ are coupled
antiferromagnetically with the isotropic exchange coupling constant $J = 25.5$\,K. Despite showing typical signatures of the 1D AFM behavior in the
static susceptibility and ESR at elevated temperatures, the Ni-hybrid compound orders AFM at $T_{\rm N} \approx 10$\,K. The occurrence of the
magnetically ordered ground state and not of the expected Haldane spin-singlet state might be presumably related to the presence of residual
interchain magnetic couplings. Still, the ESR \vl{and NMR} data indicate a possible competition between these two different states which could be
speculatively interpreted in terms of the spatially inhomogeneous ground state with coexisting AFM order and the Haldane state in the same sample. We
speculate that such an inhomogenous ground state, that was observed before in some spin-diluted Haldane magnets, could be a consequence of a small
structural disorder that promotes coupled AFM-ordered clusters around the defects in the spin chains which are intertwined with the chain segments
still exhibiting a Haldane gap.

\section{Experimental details}

\begin{figure} \centering
  \includegraphics[width=\linewidth]{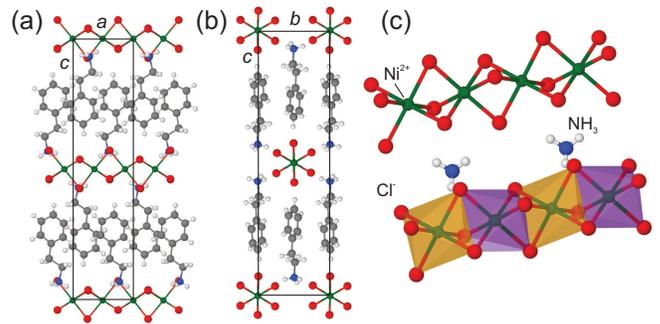}\\
  \caption[Crystal structure Ni-hybrid]{Crystallographic structure of the Ni-hybrid NiCl$_3$C$_6$H$_5$CH$_2$CH$_2$NH$_3$.
  Shown are the view on the $ac$-plane (a), $bc$-plane (b) and a close-up (c) of two chains where the face-sharing octahedra are highlighted.}
  \label{fig:Ni-structure}
\end{figure}

Samples of the Ni-hybrid compound NiCl$_3$C$_6$H$_5$CH$_2$CH$_2$NH$_3$ were grown in ethanol solution and consist of an anorganic backbone of Ni
atoms in the octahedral environment of six chlorine atoms. The general synthesis procedure and primary characterizations are described in
\cite{ArkenboutThesis2010}. \vl{The Ni-Ni distance can be subtly varied by choosing different organic constituents \cite{PolyakovThesis2015}}. The
crystallographic structure is shown in Fig.~\ref{fig:Ni-structure}. An almost perfectly symmetric octahedron is realized, with the angle between
Cl-Ni-Cl found to be $\beta_{1}\approx86^{\circ}$ and $\beta_{2}\approx94^{\circ}$. For Ni-Cl-Ni the angle is about $\gamma\approx75^{\circ}$. Along
the $c$-direction the individual Ni-chains are separated by a large organic complex consisting of a benzene structure with an amino group connected
to it by two carbon atoms. In the $b$-direction the NiCl-octahedra are separated directly through hydrogen bonds between chloride and the amino
group.

Static magnetization was measured with a VSM-SQUID magnetometer from Quantum Design Inc. which allows measurements in the temperature range from
$1.8\,\text{K}$ to $325\,\text{K}$, in magnetic fields up to 7\,T. ESR measurements with a microwave frequency of $9.6\,\text{GHz}$ and fields up to
$0.9\,\text{T}$ were performed using a standard Bruker EMX X-Band spectrometer. It is equipped with an ESR 900 He-flow-cryostat from Oxford
Instruments, which allows measurements at variable temperatures between $3.6\,\text{K}$ and $300\,\text{K}$. High-field/high-frequency ESR (HF-ESR)
was measured using a homemade spectrometer which is described in detail elsewhere \cite{Golze2006}. In the latter set-up a superconducting magnet
from Oxford Instruments can generate static magnetic fields up to $16\,\text{T}$ while a variable temperature insert enables measurements between
$1.6\,\text{K}$ and $300\,\text{K}$. For the generation and detection of microwaves with frequencies up to $360\,\text{GHz}$ a vector network
analyzer from ABmm was used. ESR measurements were performed with resonant cavities at frequencies of 9.6, 50, 83 and 93\,GHz on single crystals of
the Ni-hybrid and at frequencies up to 360\,GHz on a powder sample. For the measurements at 9.6\,GHz, several single crystals were aligned on a
teflon bar to increase the signal/noise ratio. \vl{$^{35}$Cl NMR experiments in a temperature range  1.5\,K $< T < $ 150\,K were performed with
conventional pulse NMR techniques using a Tecmag LapNMR spectrometer and a 16\,T field-sweep superconducting magnet from Oxford Instruments. The
polycrystalline powder was placed in a glass tube inside a Cu coil with a frequency of the resonant circuits of 41\,MHz.  The spectra were collected
by point-by-point sweeping of the magnetic field and integration of the Hahn spin echo at each field step. The nuclear spin-lattice relaxation rate
was measured with the saturation recovery method.}

\section{Experimental results and discussion}

\subsection{Susceptibility and Magnetization}

\begin{figure} \centering
  \includegraphics[width=0.8\linewidth]{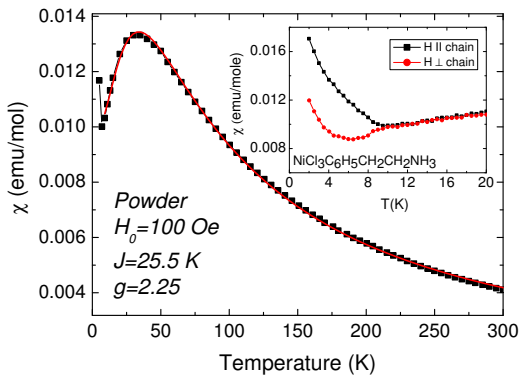}
  \caption[Ni-hybrid: Susceptibility vs. temperature]{Static susceptibility as a function of temperature.
  A broad maximum around $T\approx25$\,K is visible. Inset shows the low temperature susceptibility measured on single crystals
  with magnetic field parallel and perpendicular to the chain direction.}\label{fig:Ni-susceptibility}
\end{figure}
\begin{figure} \centering
  \includegraphics[width=0.8\linewidth]{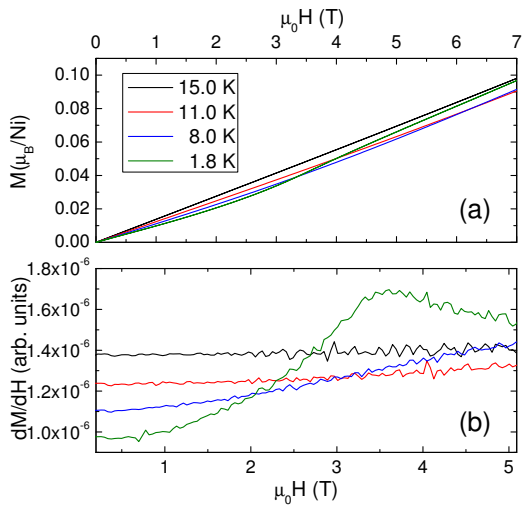}
  \caption[Ni-hybrid: Magnetization vs. magnetic field]{Magnetic field dependence of the magnetization (a)
  and its derivative $dM/dH$ (b) on powder sample. For $\mathrm{T}=15$\,K a linear increase
  in the magnetization with magnetic field is observed. At the lowest temperature a spin-reorientation is visible.}
  \label{fig:Ni-magnetization}
\end{figure}

Static susceptibility $\chi$ of the powder Ni-hybrid sample as a function of temperature at an external magnetic field of 0.01\,T is shown in
Fig.~\ref{fig:Ni-susceptibility}. $\chi(T)$ increases with decreasing temperature and shows a broad peak with a maximum around 30\,K. The
susceptibility then decreases down to about 10\,K. This is the expected behavior for a one-dimensional spin system. Indeed, as can be seen in
Fig.~\ref{fig:Ni-structure}, Ni atoms enclosed in an octahedron of oxygen or chlorine are usually in the Ni$^{2+}$ $3d^{8}$ configuration with an
effective spin moment of $S=1$ \cite{Abragam1970}. The intra-chain coupling between Ni(II)-ions is mediated by the surrounding Cl atoms of the face
sharing octahedra. The angle of $\gamma\approx75^{\circ}$ indicates an overlap of the Cl orbitals, thus an AFM superexchange is expected for the Ni
ions along the chain. Thus, from the structural point of view alone, this system seems to be a promising candidate for a Haldane system.

For a magnetically isotropic 1D system with AFM coupling the Weng equation \cite{WengThesis1968} for the susceptibility of isotropic $\mathrm{S}=1$
ring systems can be used to fit the temperature dependence of the static susceptibility \cite{Ribas1995}:
\begin{equation}\label{eq:Weng}
\chi_{S=1}=\frac{N\beta^2 g^{2}}{k_{B}T}\cdot\frac{2+0.019\alpha+0.777\alpha^2}{3+4.346\alpha+3.232\alpha^2+5.834\alpha^3}
\end{equation}
with $\alpha=J/(k_{\rm B}T)$, $k_{\rm B}$ the Boltzmann constant,  and N the number of spins. For the fit to the static susceptibility data
(Fig.~\ref{fig:Ni-susceptibility}), the $g$-factor was kept fixed with $g=2.25$ (see ESR results below) and no temperature independent offset
$\chi_{0}$ was assumed. The equation reproduces well the static susceptibility. From the fit an exchange constant of $J=25.5$\,K is extracted. For a
Heisenberg spin chain with integer spin moment $S=1$ a Haldane gap system is predicted. In the absence of single ion anisotropy the Haldane gap is
maximal and can be calculated from the exchange constant as $\Delta_{\rm H}=0.411\cdot J$ \cite{Renard2001}. With $J=25.5$\,K a Haldane gap of
$\Delta_{\rm H}=10.5$\,K is expected.

In a Haldane system the ground state is a non-magnetic singlet state. Therefore the susceptibility should go to zero with decreasing temperature. The
static susceptibility indeed decreases down to temperatures around 10\,K, but below that temperature a minimum is visible followed by an increase in
the static susceptibility with decreasing temperatures (Fig.~\ref{fig:Ni-susceptibility}).

Susceptibility measurements on single crystals reveal that the $\chi(T)$ is isotropic down to about 10\,K. Below that temperature, the static
susceptibility shows a minimum and an anisotropic increase, different for the magnetic field applied along the chain direction and perpendicular to
it (Fig. \ref{fig:Ni-susceptibility}, inset). A Curie-like increase in the susceptibility is often associated with paramagnetic impurities present in
the sample. Susceptibility from impurities could dominate over the vanishing susceptibility of a Haldane system. However, this cannot explain the
anisotropy observed.

The magnetization as a function of applied magnetic field was determined at temperatures of 1.8, 8, 11 and 15\,K up to 7\,T
(Fig.~\ref{fig:Ni-magnetization}). While at a temperature of 15\,K the magnetization increases linearly with the applied field, at 1.8\,K it shows a
non-linear behavior with an inflection point at about 3.5\,T. This is clearly visible in the derivative of the magnetization in
Fig.~\ref{fig:Ni-magnetization}(b). For the intermediate temperatures deviations from the linear increase can already be observed, but the effect is
drastically reduced. Such an inflection is usually associated with a spin-flop transition of a magnetically  ordered antiferromagnetic system, i.e. a
reorientation of spins in the increasing external field. This points to an AFM ordering.

An AFM ordering is also consistent with the susceptibility data at low temperatures. For the easy axis of an antiferromagnet the susceptibility
should go to zero, while for the easy plane it should stay constant with decreasing temperature. The susceptibility measured on the single crystals
can be interpreted as a sum of that of an antiferromagnetic state and that of paramagnetic impurities. The direction perpendicular to the chain is
the easy axis of the system. Note here that the Haldane system Pb(Ni$_{\rm 1-x}$Mg$_{\rm x})_2$V$_2$O$_8$ orders antiferromagnetically, upon
substitutional doping of Ni with Mg (S=0 impurities). In contrast to the investigations in this work, the Curie tail observed in the susceptibility
of doped PbNi$_{2}$V$_{2}$O$_{8}$ is suppressed at the onset of AFM order \cite{Uchiyama1999}. This indicates that in the Ni-hybrid compound (not
all) impurities are involved in the magnetic ordering.

The total magnetization is quite small with $M \approx0.1\mu_{\rm B}/\mathrm{Ni}$ at the maximum field of 7\,T. For the Ni$^{2+}$ ions contributing
to the magnetization a saturation field of $M_{sat}= gS=2.25$\,$(\mu_{\rm B}/\mathrm{Ni})$ is expected. This is consistent with one-dimensional
Heisenberg AFM as well as Haldane systems, in which the saturation magnetization can often not be reached even in fields up to 40\,T
\cite{Uchiyama1999,Masuda2002}.

Altogether, the results of static susceptibility and magnetization on the Ni-hybrid samples indicate a one-dimensional spin chain which exhibits an
antiferromagnetically ordered ground state that develops below $T_{\rm N} \approx 10$\,K  with the easy axis perpendicular to the chain direction and
a spin-flop transition at $H_{\rm c} = 3.5$\,T. A certain amount of impurities is present in the sample.

\subsection{Electron Spin Resonance}
\begin{figure}
\centering
  \includegraphics[width=0.8\linewidth]{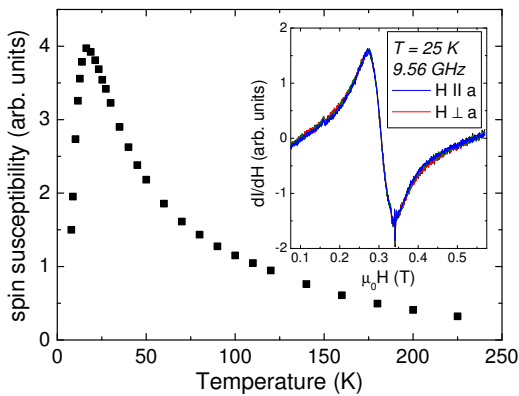}
  \caption[Ni-hybrid: Spin-susceptibility]{Temperature dependence of the spin susceptibility
   as determined from ESR measurements around 9.56\,GHz. Inset shows the ESR-spectra (absorption derivative) at $T=25$\,K
   for magnetic fields applied parallel and perpendicular to the chain direction.}\label{fig:Ni-spin-susceptibility}
\end{figure}
To get a more detailed picture about the physics of the Ni-hybrid spin chain compound ESR measurements were conducted. ESR is a valuable tool to
probe the local static and dynamic magnetic properties which also can give information about the different energy states in one-dimensional
Heisenberg antiferromagnets \cite{Sieling1995,Hagiwara1990,Affleck1992,Smirnov2002, Smirnov2008,Krug2009}.
%
%ESR measurements were performed with resonant cavities at frequencies of 9.6, 50, 83 and 93\,GHz on single crystals of the Ni-hybrid and at
%frequencies up to 360\,GHz on a powder sample. For the measurements at 9.6\,GHz, several single crystals were aligned on a teflon bar to increase the
%signal/noise ratio.
%
%
%
A single-crystalline ESR spectrum at 9.6\,GHz (X-Band) at $T=25$\,K is shown in Fig.~\ref{fig:Ni-spin-susceptibility} (inset) for external magnetic
fields applied along the chain direction and perpendicular to it. The resonance signal exhibits a Lorentzian line around a resonance field of about
0.3\,T corresponding to a $g$-factor of $g=2.25$. This $g$-factor is typical for a Ni$^{2+}$ ion in an octahedral crystal field \cite{Abragam1970}.
The ESR signals are almost perfectly isotropic over the whole temperature range down to about 8\,K. This indicates the absence of single ion
anisotropy [$D=0$ in Eq.~(\ref{Hamilton})] in the Ni-hybrid compound meaning that an isotropic Heisenberg spin chain is realized in the paramagnetic
state which is very favorable for the realization of a Haldane system. The absence of single ion anisotropy is most likely related to the regular
octahedral ligand coordination of Ni$^{2+}$. In such high symmetry of the ligand crystal field the $S=1$ state of Ni$^{2+}$ remains 3-fold degenerate
in zero magnetic field implying the isotropic character of the Ni spin \cite{Abragam1970}. The integrated intensity $I_{\rm ESR}$ of an ESR signal is
determined by the intrinsic susceptibility $\chi_{\rm spin}$ of the spins participating in the resonance \cite{Abragam1970}. From Lorentzian fits of
the ESR signals of the Ni-hybrid sample $I_{\rm ESR} \propto \chi_{\rm spin}$ can be evaluated and the temperature dependence of $\chi_{\rm spin}$ is
plotted in Fig.~\ref{fig:Ni-spin-susceptibility}. The spin susceptibility shows a maximum around 20\,K with a sharp decrease at the low temperature
side. This decrease is associated with the broadening and the strong decrease in the amplitude of the resonance. It is clearly visible that, in
comparison to the bulk static susceptibility (Fig.~\ref{fig:Ni-susceptibility}), the spin susceptibility exhibits an enhanced maximum and a steeper
decrease when going to lower temperatures. From Fig.~\ref{fig:Ni-spin-susceptibility} it is apparent that towards zero temperature $\chi_{\rm spin}$
would approach the zero value. Below 8\,K the signal cannot be observed at this frequency. This clear tendency towards zero spin susceptibility
indicates that the Ni spin system (giving rise to this spectrum) seems to behave as a Haldane system down to 8\,K. This is in contrast to the
observed magnetic order at $T_{\rm N} \approx 10$\,K in the static susceptibility and magnetization measurements.

However, there can be other spin subsystems present which cannot be observed by ESR at frequencies around 10\,GHz. Especially ESR resonances
associated with the AFM ordering which was detected by susceptibility and magnetization measurements may occur outside of the field and the frequency
ranges of the X-Band spectrometer. Thus, HF-ESR experiments at higher frequencies and higher magnetic fields were performed as well.
\begin{figure}
 \centering
  \includegraphics[width=\linewidth]{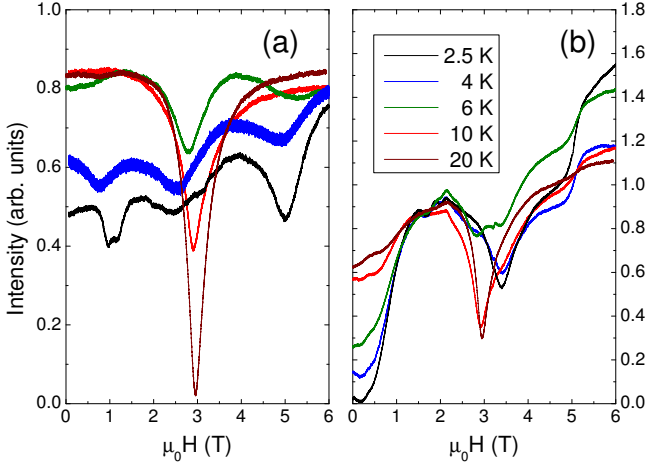}
  \caption[Ni-hybrid: HF-ESR spectra at 93\,GHz]{ESR spectra at 93\,GHz with magnetic field perpendicular to the chain (a) and along the chain
  (b) for temperatures from 20\,K down to 2.5\,K. While the central line decreases additional lines develop with decreasing
  temperature.}\label{fig:Ni93GHz}
\end{figure}

Similar to the X-band results, at all higher probing frequencies up to 360\,GHz a single isotropic line is observed for temperatures above 10\,K.
Fig.~\ref{fig:Ni93GHz} shows the ESR spectra at a frequency of about 93\,GHz for selected temperatures between 20\,K and 2.5\,K measured on one
single crystal. ESR spectra with the magnetic field perpendicular to the chain direction are shown in Fig.~\ref{fig:Ni93GHz}(a). At 20\,K only a
single Lorentzian line is visible, in agreement with the experiments at 10\,GHz. With decreasing temperature this line decreases in intensity and
shifts to lower fields. Interestingly, two additional lines appear at fields of about 1\,T and 5\,T and become stronger in intensity with decreasing
temperatures. This is a new feature not observed at lower frequencies.

For the magnetic field parallel to the chain [Fig.~\ref{fig:Ni93GHz}(b)], the central line also decreases, but shifts to higher fields. At almost
zero magnetic field another signal appears. However it is not clear if the minimum of the spectrum is fully visible. That is why the absolute value
of the resonance field extracted cannot be very accurate. At around 5\,T another feature is observed which appears phase-shifted with respect to the
central line. This probably does not originate from the main crystal and could be a spurious effect due to some fragment at another position in the
resonator.

\begin{figure}
 \centering
  \includegraphics[width=0.8\linewidth]{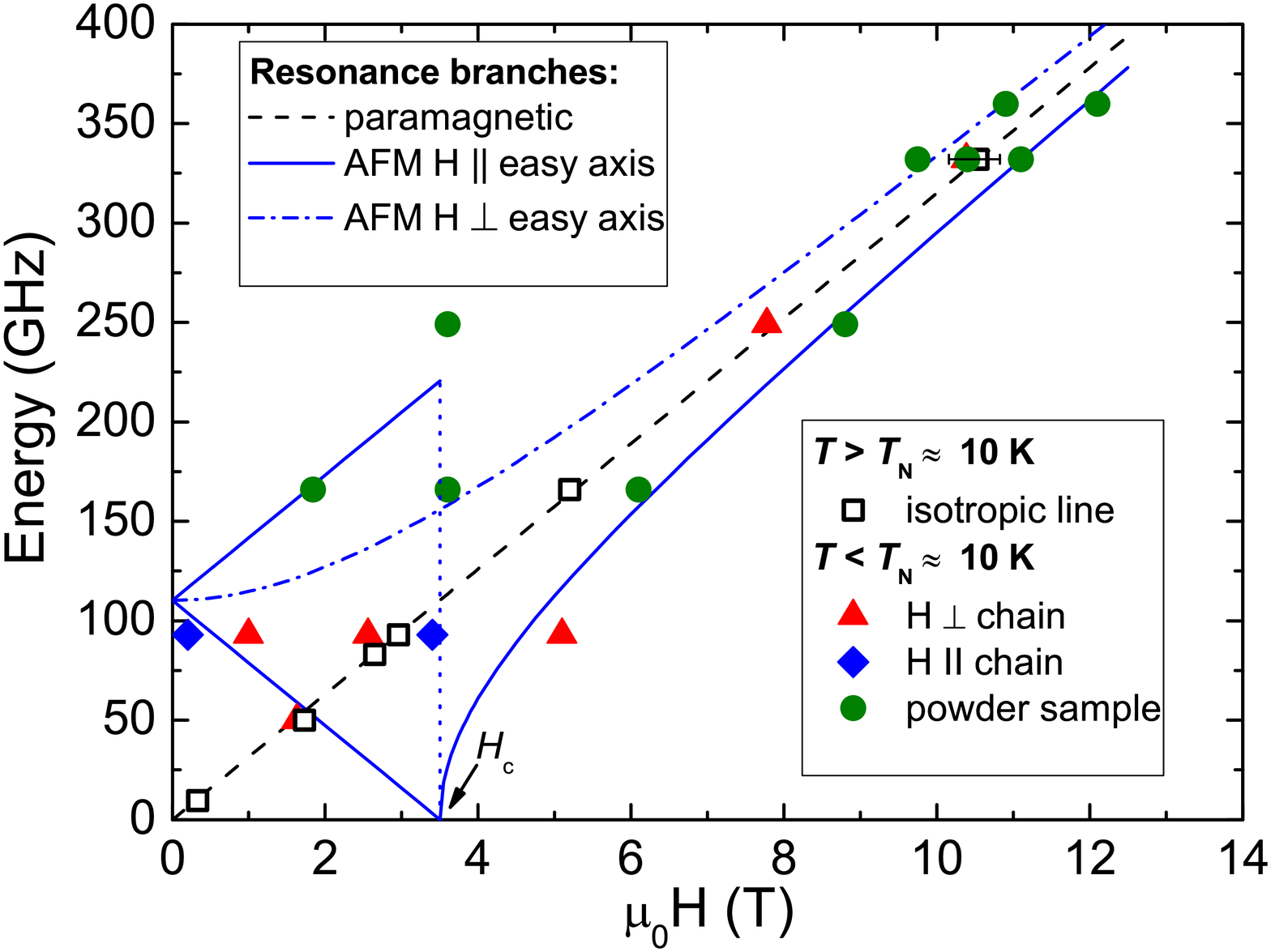}
\caption[Ni-hybrid: Field dependence of ESR spectra]{Summary of the ESR modes in the frequency vs. magnetic field plot. Open squares correspond to
the isotropic ESR signals observed in the paramagnetic regime at $T > 10$\,K. Triangles and diamonds denote the ESR modes detected at $T < T_{\rm N}
\approx 10$\,K for the external field applied perpendicular and parallel to the Ni-chain axis, respectively. Circles depict the signals of a powder
sample at $T < T_{\rm N} \approx 10$\,K. Dash line corresponds to the paramagnetic branch $\nu = (g\mu_{\rm B}/h)H$, and the solid and dash-dot
curves represent the AFM branches for the easy- and the hard directions of a collinear two-sublattice antiferromagnet according to
Eqs.~(\ref{easy1},\ref{easy2}) and Eq.~(\ref{hard1}), respectively. $H_{\rm c}$ denotes the spin-flop field determined by the magnetization
measurements. Note that the signals grouped around the paramagnetic branch strongly decrease in intensity below $\sim 20$\,K, whereas the signals
grouped around the AFM branches appear first below $T_{\rm N} \approx 10$\,K and grow in intensity at lower temperatures.}\label{fig:Ni-fvsH}
\end{figure}

The above discussed ESR signals from single crystals of the Ni-hybrid compound measured with the resonator-based setups at $10 - 93$\,GHz as well as
data for a powder sample measured without resonators at higher frequencies up to 360\,GHz are summarized in a frequency $\nu$ vs. magnetic field $H$
chart in Fig.~\ref{fig:Ni-fvsH}. The paramagnetic signals at temperatures above 10\,K follow a linear dependence $\nu = (g\mu_{\rm B}/h)H$ (dashed
line) with the slope given by the $g$-factor $g=2.25$. Here $h$ is the Plank constant. The theoretical equations \cite{Turov1965,Foner1963} for
resonances of a collinear two-sublattice antiferromagnet are shown by solid curves (for $H\parallel$ easy axis) and by a dash-dot line (for $H\perp$
easy
axis):\\
\\
$H \parallel$ easy axis, $H<H_{c}$:
\begin{equation}
\nu_{1,2}=\Delta_{\rm a}\pm\frac{g\mu_{\rm B}}{h} H \label{easy1}
\end{equation}
$H \parallel$ easy axis, $H>H_{c}$:
\begin{eqnarray}
\nu_{1}=0&      &\nu_{2}=\sqrt{\left(\frac{g\mu_{\rm B}}{h} H\right)^2-\Delta_{\rm a}^2} \label{easy2}
\end{eqnarray}
$H \perp$ easy axis:
\begin{eqnarray}
&\nu & =  \sqrt{\left(\frac{g\mu_{\rm B}}{h} H \right)^2+\Delta_{\rm a}^2}, \label{hard1}
\end{eqnarray}
where $g=2.25$ and the so-called magnetic anisotropy gap at zero field $\Delta_{\rm a}=(g\mu_{\rm B}/h)H_{\rm c}$ with $\mu_0H_{\rm c}=3.5$\,T being
the spin-flop magnetic field value obtained from magnetization measurements.

As known from the susceptibility measurements on the single crystal the easy axis is perpendicular to the chain. The side peaks at 93\,GHz for this
direction [Fig.~\ref{fig:Ni93GHz}(a)] roughly agree with the theoretical AFM resonance modes (Fig.~\ref{fig:Ni-fvsH}). A similar assignment can be
made for the ESR signals for $H\parallel$ chain axis as the hard-direction AFM modes. For a powder sample, modes for both directions are present due
to the powder averaging. Notably, the signals at all measured frequencies that follow the paramagnetic resonance branch lose their intensity below
$\sim 20$\,K as if they corresponded to some excited magnetic state that gets thermally less populated with decreasing temperature. These signals are
still detectable below $T_{\rm N} \approx 10$\,K where their position shifts as if the resonating spins sensed the internal magnetic fields caused by
the antiferromagnetic order in their vicinity.

Based on the ESR results, one could conjecture that in the studied Ni-hybrid compound two spin subsystems could be realized, the one which orders AFM
at $T_{\rm N} \approx 10$\,K, and the other one which shows signatures of thermally activated paramagnetism. It is tempting to speculate that the
latter subsystem might develop the Haldane spin gap and could be spatially separated but yet still coupled to the AFM ordered subsystem.

Indeed, a coexistence between singlet quantum ground state and classically ordered magnetic state was reported, e.g.,  for the Haldane compound
PbNi$_2$V$_2$O$_8$ where the spinless defects were introduced in the Ni-spin chain by Mg doping \cite{Uchiyama1999,Smirnov2002}. The development of
the AFM order was attributed to the nucleation of the soliton-like AFM clusters around the defect sites in the Haldane chain which couple together
due to residual interchain magnetic exchange. ESR experiments have indicated that at small concentration of defects Pb(Ni$_{\rm 1-x}$Mg$_{\rm
x})_2$V$_2$O$_8$ ($x \leq 0.02$) develops a spatially inhomogeneous state of co-existing large AFM ordered clusters, small paramagnetic clusters, and
spin-singlet Haldane regions \cite{Smirnov2002}.

The Ni-hybrid compound studied in the present work was not doped intentionally with nonmagnetic defects. However, it is conceivable that there might
be some (small) structural disorder in the crystals, resulting in a segmentation of the Ni-chains in fragments of different length. At the chain ends
uncompensated spins and/or AFM correlated regions could develop and interact with each other, and be responsible for the small Curie-like upturns of
the static susceptibility at low temperatures and AFM order at $T_{\rm N} \approx 10$\,K, as evidenced by the static magnetic and ESR measurements.
On the other hand, one cannot completely exclude the possibility that still a certain amount of Ni-chains in the sample develop the Haldane
spin-singlet ground state which could explain the thermally activated paramagnetic ESR signals which are detected on the background of the AFM
resonance modes.

Finally, it should be noted that ESR spectra were observed in Haldane systems, which were attributed to singlet-triplet transitions between the $S=0$
ground state and the $S=1$ triplet state in NENP \cite{Date1990,Affleck1992} and PbNi$_2$V$_2$O$_8$ \cite{Smirnov2008}. These transitions are
forbidden by the dipole selection rules. However, mixing between pure $S = 0$ and $S = 1$ spin states is possible through anisotropic exchange
interactions or single ion anisotropy. Then the forbidden transitions can be observed in an ESR experiment. For the Ni-hybrid compound ESR data in
the paramagnetic state evidence the absence of single ion anisotropy ($D = 0$). Therefore it is likely that the mixing is too small to make an
observation of the forbidden transitions possible.

\subsection{NMR spectroscopy}

\begin{figure}
\centering
  \includegraphics[width=1.0\linewidth]{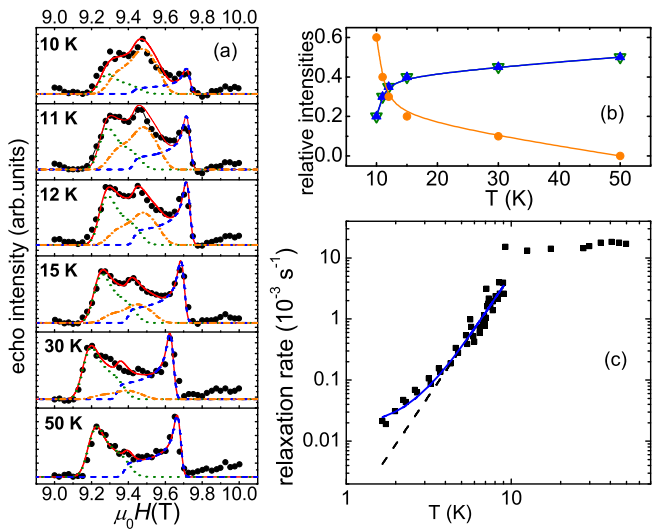}
  \caption{\vl{(a) Selected $^{35}$Cl NMR spectra (solid circles) in the temperature range  10\,K $< T <$ 50\,K. Solid line represents a modelling of the
three-component powder averaged spectra, dash, dot and dash-dot lines show different spectral contributions (see the text); (b) Temperature
dependence of the relative intensities of the central spectral contribution (solid circles) and of the left and right side spectral contributions
[almost coinciding up- (blue) and down (green) triangles]; (c) Temperature dependence of the $^{35}$Cl nuclear spin-lattice relaxation $T_1^{-1}$
(solid squares). Dash line represents a fit by the power law $\sim T^{3.89}$, solid line shows a sum of the power law $\sim T^{4.13}$ and of the
activation law $\sim \exp(-3.2/T)$ (see the text).}}
   \label{NMR}
\end{figure}

\vl{Additional insights into the local magnetic properties and the spin dynamics of the Ni-hybrid compound were obtained by $^{35}$Cl NMR
spectroscopy. The NMR spectrum has a total width of more than 2\,T and consists of two structured peaks corresponding to two isotopes of  $^{35}$Cl
and $^{37}$Cl and a quadrupole background. The temperature evolution of the main line for $^{35}$Cl nuclei below a temperature of  50\,K is shown in
Fig.~\ref{NMR}(a). The "two-horn" shape of the spectrum is present approximately down to 30\,K, transforming with a further temperature decrease into
the three peaks structure where the additional central line shifts gradually to higher fields towards the Larmor field. The second isotope $^{37}$Cl
line undergoes the same changes. Below 10\,K, the shape of the spectrum changes dramatically: the signals identified above as the main lines for both
Cl isotopes disappear, and the total width of the spectrum increases by approximately 1\,T already at $T = 8$\,K. Such a sharp transformation of the
NMR spectrum suggests the occurrence of the antiferromagnetic transition around 9\,K, consistent with the N\'{e}el temperature $T_{\rm N} \approx
10$\,K determined by the static magnetization measurements. The temperature dependence of the nuclear relaxation rate $T_1^{-1}$ [Fig.~\ref{NMR}(c)]
measured at the midpoint of the spectrum exhibits a broad maximum around 35\,K similar to the behavior of the static susceptibility. At $T \approx
10$\,K the temperature dependence has a weak maximum while with a further temperature decrease the relaxation rate drops sharply. A sharp change in
the temperature dependence at 10\,K indicates a possible phase transition at this temperature, again in agreement with the static data, while the
absence of a pronounced maximum, which is typical for establishing the magnetic order in 3D systems, suggests the magnetic quasi-one-dimensionality
of the Ni-hybrid compound.}

\vl{The NMR measurements were performed in fields of the order of 9\,T. In such fields, the hypothetical Haldane gap, which value at $H = 0$ is
estimated theoretically as 10.5\,K, is expected to be almost closed due to the lowering of the energy of the $S_z = |-1\rangle$ state of the excited
$S = 1$ triplet. We attempted to describe the transformation of the powder spectra near the N\'{e}el temperature as a result of the change of the
symmetry of the local fields acting on the Cl nuclei due to the development of the Ni spin correlations in the chain. Though no perfect fit to the
experimental spectra could be achieved, the reasonable agreement between the model and experiment requires an assumption of the appearance and growth
of an additional component with a different symmetry which is located between the components of the high-temperature spectrum [Fig.~\ref{NMR}(a)].
The appearance of this signal can be speculatively explained assuming a phase separation in the chains into the regions where the growth of AFM
correlations leads to the ordering and the regions where the Haldane nonmagnetic state could still develop with decreasing temperature. The magnitude
and symmetry of the field at the Cl nuclei in the "Haldane-like" part of the chains is determined by nearest magnetic regions, and the intensity of
this NMR signal increases with decreasing temperature. Concomitantly,  the intensity of the left and right signals originating from the paramagnetic
regions falls down [Fig.~\ref{NMR}(b)] despite increasing their width. Another possible origin of this mid-signal in the NMR spectrum could be the Cl
nuclei near the chain defects, the response from which becomes more pronounced when the Ni spins in the chains begin to correlate. Also the behavior
of the relaxation rate $T_1^{-1}(T)$ suggests a situation more complex than a simple AFM ordering model. In the AFM ordered state, $T_1^{-1}(T)$ is
mainly driven by magnon scattering, leading to a power-law temperature dependence \cite{Belesi2006,Yogi2015}. In the limit where the temperature is
much higher than the anisotropy gap in the spin-wave spectrum, $T_1^{-1}(T)$ either follows a $T^3$ dependence due to a two-magnon Raman process or a
$T^5$ dependence due to a three-magnon process. If the temperature is smaller than the gap, the relaxation rate is proportional to $T^2\cdot\exp
(-\Delta_{\rm a}/k_{\rm B}T)$ where $\Delta_{\rm a}$ is the spin-wave anisotropy gap. In the Ni-hybrid compound the relaxation below 10\,K does not
obey this combined power-exponential law. The temperature dependence can be well described by the $T^{3.89}$ power law down to $\sim 3$\,K suggesting
that the relaxation is mainly governed by the two-magnon process [Fig.~\ref{NMR}(c)]. However, in order to fit also the low-temperature part of the
$T_1^{-1}(T)$ dependence, we need to add the second term, which has an activation character with a gap of about 3.2\,K [Fig.~\ref{NMR}(c)]. In this
case the first term is proportional to $T^{4.13}$ suggesting that the relaxation is likely governed mainly by the three-magnon process. If this gap
is a conjectured Haldane gap, its value of 3.2\,K would be presumably a bit too large for such a high field of the measurement. Nevertheless, within
the speculative phase separation scenario one would indeed expect that just below $T_{\rm N}$ the nuclear spin relaxation is determined by regions
with a N\'{e}el order. As the temperature is lowered, this relaxation channel strongly slows down and the contribution from the "Haldane-like"
regions becomes noticeable.}
\\

\section{Conclusions}

In summary, we have studied magnetic properties of the Ni-hybrid compound NiCl$_3$C$_6$H$_5$CH$_2$CH$_2$NH$_3$ by static magnetic as well as ESR and
NMR measurements. In this material structurally well-defined Ni spin-1 antiferromagnetic chains are realized. ESR data in the paramagnetic state at
elevated temperatures evidence an isotropic Heisenberg character of the Ni spins. The analysis of the temperature dependence of the static
susceptibility yields an AFM intra-chain exchange interaction constant $J = 25.5$\,K. Though the above results suggest this Ni-hybrid compound as a
possible realization of the Haldane spin-1 AFM chain that should develop a singlet ground state with the quantum spin gap $\Delta_{\rm H}=0.411\cdot
J = 10.5$\,K, experimental data show that this material orders AFM at $T_{\rm N} \approx 10$\,K \vl{and the excitations below $T_{\rm N}$, as probed
by NMR, are predominantly magnon-like}. The AFM order and not the Haldane spin-singlet ground state is possibly due to the non-negligible interchain
interactions. Interestingly, besides the AFM resonance modes detected at $T < T_{\rm N}$, a thermally activated paramagnetic ESR signal is observed
in the spectra measured at different frequencies. It could be compatible with the thermally activated signal of a Haldane chain. To explain this
signal, \vl{as well as the unusual shape transformation of the $^{35}$Cl NMR spectrum and the peculiar behavior of the nuclear spin relaxation rate},
we speculate that a spatially inhomogeneous ground state might be realized in the Ni-hybrid compound. Assuming the occurrence of a small amount of
structural defects in the Ni-chain that cut the spin chain in fragments of different length one could conjecture the nucleation of the soliton-like
AFM clusters at the chain ends which couple together and order AFM. We further speculate that besides the ordered phase there might be mesoscopically
spatially separated regions in the sample where the Ni-chains still exhibit a Haldane spin gap, similar to the situation which was reported for the
spin-diluted Haldane magnet Pb(Ni$_{\rm 1-x}$Mg$_{\rm x})_2$V$_2$O$_8$.

\section{Acknowledgements}

This work was supported in parts by the Deutsche Forschungsgemeinschaft (DFG) through projects FOR912 and KA 1694/8-1, by the Dieptestrategie of the
Zernike Institute for Advanced Materials, and by the Erasmus+ ICM Program of the European Union.

%\bibliography{thesisbib2}


\end{document}